\documentstyle[12pt]{article}
\begin{document}
{\pagestyle{empty}

{\Large\begin{center}
{\bf A UNIFIED CONFORMAL MODEL \\ FOR FUNDAMENTAL
INTERACTIONS \\ WITHOUT DYNAMICAL HIGGS FIELD}
\end{center}}
\vskip 1cm

\begin{center}
{\bf Marek Paw\l owski}$^{\ast\dagger}$
\\
Soltan Institute for Nuclear Studies, Warsaw, POLAND
\\
{\bf and }
\\
{\bf Ryszard R\c aczka}$^\ast\ddagger$
\\
Soltan Institute for Nuclear Studies, Warsaw, POLAND
\\
and
\\
Interdisciplinary Laboratory for Natural and Humanistic Sciences
\\
International School for Advanced Studies (SISSA), Trieste, ITALY
\\ ~ \\
{July,$\;\;\;$1994}
\end{center}

\vskip -12cm
\rightline{ILAS 4/94}
\vskip 11cm

\bigskip\bigskip
\centerline{\bf Abstract}

\bigskip

A Higgsless model for strong, electro--weak and gravitational
interactions is proposed.  This model is based on the local symmetry
group $SU(3)\times SU(2)_{L}\times U(1)\times C$ where $C$ is the local
conformal symmetry group.  The natural minimal conformally invariant
form of total lagrangian is postulated.  It contains all Standard
Model fields and gravitational interaction.  Using the unitary gauge
and the conformal scale fixing conditions we can eliminate all four
real components of the Higgs doublet in this model.  However the
masses of vector mesons, leptons and quarks are automatically
generated and are given by the same formulas as in the conventional
Standard Model.  In this manner one gets the mass generation without
the mechanism of spontaneous symmetry breaking and without the
remaining real dynamical Higgs field.  The gravitational sector is
analyzed and it is shown that the model admits in the classical limit
the Einsteinian form of gravitational interactions.

\vfil
\vskip.5cm
\hrule\smallskip

{\footnotesize
\leftline{$^\ast$~~Partially supported by Grant No. 2 P302 189 07 of
Polish Committee for Scientific}

\leftline{Research.}

\leftline{$^{\dagger}$~~e--mail: PAWLOWSK@fuw.edu.pl}

\leftline{$^\ddagger$~~e--mail: RRACZKA@fuw.edu.pl}}

\eject
{}~

\eject}

\section{Introduction}

The recent evidence for top quark production with the top mass
estimated as $m_t=174\pm 10^{+13}_{-12} GeV$ \cite{fnal} implies that
the Higgs particle -- if exists -- may have the mass of the order of
1TeV:  in fact the central value of $m_H$ implied by the present data
of $m_t$ and $m_W$ was estimated by Hioki and Najima \cite{hioki} at
$m_H\approx 1700 GeV$ with an enormous error however.  Since in the
lowest order $\lambda={1\over2} ({m_H\over v })^2$ one can afraid that
the Higgs self-coupling $\lambda$ would be also very large
($\lambda\approx 25$ for the central value of $m_H$ given by Hioki and
Najima).  Such strong Higgs self-interaction would mean that the loops
with Higgs particles would dominate all other contributions.  Therefore
the perturbative predictions in Standard Model(SM)
for many quantities become
unreliable.  Hence the predictive power of the SM and
its consistency may be questionable 

The Higgs particle with such a large mass becomes suspicious. It
is natural therefore to search for a modification of SM in which
all confirmed by experiment particles would exist but the Higgs
particle as the observed object would be absent.

We show in this work that such a modification of SM is possible under
the condition that one joints to strong and electro--weak interactions
also the gravitational interaction.  This extension of the class of SM
interactions is in fact very natural.  Indeed whenever we have the
strong and electro--weak interactions of elementary particles, nuclea,
atoms or other objects we have also at the same time the gravitational
interactions.  It seems natural therefore to consider an unified model
for strong, electro--weak and gravitational interactions which would
describe simultaneously all four fundamental interactions.  It is well
known that gravitational interactions give a negligible effect to most
of strong or electro--weak elementary particle processes.  We show
however that they may play the crucial role in a determination of the
physical fields and their masses in the unified model and that their
presence allows to eliminate all Higgs fields from the final
lagrangian.

In turn we recall that in the conventional Standard Model the Higgs
mechanism of spontaneous symmetry breaking (SSB) provides a simple and
effective instrument for mass generation of weak gauge bosons, quarks
and leptons.  However, despite of many efforts of several groups of
experimentalists \cite{LEP} the postulated Higgs particle of the SM
was not observed.  Hence one might expect that the model for strong
and electro--weak interactions supplemented by the gravitational
interaction in which all dynamical Higgs fields may be eliminated can
provide a natural frame--work for a description of elementary particle
fundamental interactions.

In order to construct a new form of total lagrangian for
the theory of strong and
electro--weak interactions extended by the gravitational interactions
we observe that the gauge symmetry $SU(3)\times SU(2)_L \times U(1)$
of the fundamental interactions may be naturally extended by the local
conformal symmetry.  The choice of the unitary gauge condition for
$SU(2)_L$ gauge group allows to eliminate the three out of four Higgs
fields from the complex Higgs doublet.  In turn the choice of the
scale fixing condition connected with the local conformal symmetry
allows to eliminate the last Higgs field.  In that manner all four
Higgs fields can be gauged away completely!  It is remarkable that in
spite of the elimination of all Higgs fields in our model the vector
meson, lepton and quark masses are generated and at the tree level
they are given by the same analytical formulas as in the conventional
SM.

Thus it may be that the dynamical real Higgs field and the associated
Higgs particles are in fact absent and it is therefore not surprising
that they could not be detected in various experiments \cite{LEP}.

We review in Section 2 the present problems with a very massive Higgs
particle.  Next in Section 3 we discuss the properties of local
conformal symmetry and its representations in field space of arbitrary
spin.  We present in Section 4 the form of the total lagrangian of our
unified theory of electro--weak, strong and gravitational interactions
determined by the gauge and the local conformal invariance.  The
noteworthy feature of the obtained lagrangian is the lack of the Higgs
mass term $\mu^2 \Phi^\dagger\Phi$.  We show next that using the
unitary gauge condition and the conformal scale fixing condition we
can eliminate all dynamical Higgs fields from the theory!  We show in
Section 5 that in spite of the lack of dynamical Higgs fields the
masses of vector mesons, leptons and quarks are generated and at the
tree level are given by the same analytical expressions in terms of
coupling constants as in the conventional SM.

We stress that the renormalizability of our model depends on the value
of the new coupling constant $\beta$ which determines the properties
of gravitational sector.  We discuss in Section 6 the variant of our
model with $\beta\not=0$.  This leads to the model with massive vector
mesons which is nonrenormalizable.  In order to get definite
perturbative predictions -- especially for electro--weak processes --
we have to introduce the ultraviolet cutoff $\Lambda$.  We show the
close connection between the large Higgs mass $m_H$ and $\Lambda$.  We
illustrate this relation in the case of universal electro--weak
parameters $\varepsilon_{N1}$, $\varepsilon_{N2}$ and
$\varepsilon_{N3}$ of Altarelli {\sl et al.}  \cite{altarelli} for
which we show that the difference between SM results for
$\varepsilon_{Ni}$ and in our model is essentially proportional to
$\log {\Lambda^2\over m_H^2}$; thus if one chooses $\Lambda\cong m_H$
one obtains the same analytical formulas for $\varepsilon_{Ni}$ in SM
and in our Higgsless model.  We show also how using so called General
Equivalence Theorems one can calculate the high energy limit for
various processes in our model.

We present in Section 7 the analysis of the gravitational sector in
the unified model.  We show that our unified model after determination
of the unitary gauge and scale fixing leads already at classical level
to the conventional gravitational theory with Einstein--Hilbert
lagrangian implied by the conformal Penrose term contained in the
unified lagrangian.

We present in Section 8 the special version of our model (with
$\beta=0$) which may lead to perturbatively renormalizable model of
fundamental interactions.  We discuss shortly some open problems of
this formulation of the unified theory.

Finally we discuss in Section 9 three alternatives for a description
of fundamental interactions which are given by the conventional SM or
its extensions, Higgsless renormalizable SM and nonrenormalizable
Higgsless models.  We discuss also some open problems connected with
derivation of predictions in low and high energy regions from
nonrenormalizable Higgsless models.

The present work is the extension of our previous paper \cite{babbage}
and contains the answer to several questions raised by its readers.

\section{Difficulties with Standard Model Higgs particle.}

We shall argue that the recently announced \cite{fnal} evidence for
the top quark with the mass

$$ m_t=174\pm 10^{+13}_{-12} GeV \eqno (2.1)$$
may lead to a serious conceptual and calculational problems in the
Standard Model.  The relatively heavy top quark with the mass (2.1) --
heavier than expected on the base of LEP1-CDF-UA1 data \cite{elis},
\cite{oldaltarelli} -- shifts up the expected region of SM Higgs mass
and consequently also the area of expected Higgs quartic self-coupling
$\lambda$.  The analysis of the value of Higgs mass following from the
one-loop formula for the $W$--meson mass carried out by Hioki and
Najima \cite{hioki} leads to the central value
$$m_H\cong1700GeV. \eqno (2.2) $$

Since the Higgs self--coupling constant $\lambda$ and the Higgs mass
are connected at the tree level by the formula

$$\lambda ={1\over 2} ({m_H\over{<\phi >}})^2, \hskip1cm {<\phi
>}=246GeV \eqno (2.3)$$
the value (2.2) implies that

$$ \lambda\cong 25. \eqno (2.4)$$

This looks very dangerous; however to be honest we should mention that
within the present experimental errors for $m_t$ given by FNAL result
and for other experimental quantities being the input for the
estimation (2.2) there is a considerable admissible deviation for
$m_H$ from the central value $1700GeV$
\cite{passarino}\cite{pietnascie}.  Consequently $m_H$ and therefore
also $\lambda$ may be much smaller.

Despite the fact that the present electro-weak data are not very
conclusive the result (2.1) compels many authors to consider the
possibilities of large Higgs mass and strong Higgs self--coupling more
seriously.  The super--strong Higgs self--coupling (like (2.4) or even
smaller) would evidently break--down the perturbative calculations for
many processes for which Higgs loops with $\lambda$-coupling
contributes.  For instance the two-loop perturbation expansion for the
partial width decay $\Gamma(H\rightarrow\bar{f}f)$ of the Higgs
particle into the fermion -- antifermion pair can be written in the
form

$$\Gamma(H\rightarrow\bar{f}f)=\Gamma_0[1+0.11({m_H\over 1TeV})^2 -
0.78({m_H\over 1TeV})^4] \eqno (2.5)$$
where $\Gamma_0$ is the partial width in the Born approximation and
the second and third term in the bracket represent the one- and the
two-loop contributions respectively \cite{knie}.

We see that with increasing $m_H$ the importance of the two-loop
contribution rapidly increases:  for $m_H>375GeV$ the two-loop
contribution dominates the one-loop and for $m_H>1200GeV$ the width
becomes negative!  This demonstrates the complete breakdown of
perturbation theory for the Higgs mass of the order of 1TeV.

We see therefore that the supposition that the real Higgs field and
the corresponding Higgs particle exists in the SM may lead
to rather fundamental conceptual and calculational difficulties.
Therefore it seems justified at present to look for a modification of
SM in which all experimentally confirmed facts would be reproduced but
the Higgs particle as the observed object would not exist.

Recently there were proposed several Higgsless models for
electro--weak and strong interactions.  In particular Shildknecht and
collaborators proposed the Higgsless massive vector boson model
\cite{schild} and they have compared  some of its predictions
with the predictions of the conventional SM.  In the work
\cite{dipdip} it was proposed a Higgsless SM with nonrenormalizable
current--current and dipol--dipol interactions.  Finally in
\cite{sigmaSM} it was proposed a gauged $\sigma$--model for
electro--weak interactions.

It seems to us that our Higgsless model based on the extension of
electro--weak and strong interactions by gravitational interactions
which leads to the extension of gauge symmetry by the local conformal
symmetry presents a most natural frame--work for fundamental
interactions.

\section{Local conformal symmetry}

Let $M^{3,1}$ be the pseudo--Riemannian space time with the metric
$g_{{\alpha \beta}}$ with the signature $(+,-,-,-)$.  Let $\Omega(x)$
be a smooth strictly positive function on $M^{3,1}$.  Then the
conformal transformation in $M^{3,1}$ is defined as the transformation
which changes the metric by the formula $$g_{{\mu\nu}}(x)\rightarrow
\tilde{g}_{{\mu\nu}}(x)=\Omega^2(x) g_{{\mu\nu}}(x).  \eqno (3.1)$$
The set of all conformal transformations forms the multiplicative
abelian infinite--dimensional group $C$ with the obvious group
multiplication law.

It is evident from (3.1) that $(M^{3,1},g_{\mu\nu})$ and
$(M^{3,1},\tilde g_{\mu\nu})$ have identical causal structure and
conversely it is easy to show that any two space times which have
identical causal structure must be related by a local conformal
transformation.

The conformal transformations occur in many problems in general
relativity.  In particular Canuto et.  al.  proposed the
scale--covariant theory of gravitation, which provides an interesting
alternative for the conventional Einstein theory \cite{canuto}.

It should be stressed that a conformal transformation is not a
diffeomorphism of space time.  The physical meaning of the conformal
transformations follows from the transformation law of the length
element
$$dl(x)= \sqrt{-g_{ij}dx^i dx^j} \hskip1cm \rightarrow \hskip1cm
\tilde{dl(x)}=\Omega(x) dl(x). \eqno (3.2)$$
Hence a local conformal transformation changes locally the length
scale.  Since in some places of the Earth one utilizes {\sl the meter}
as the length scale, whereas in other places one utilizes {\sl the
feet} or {\sl the ell} as the length scale one my say that one
utilizes the local conformal transformations in everyday live.
Similarly one verifies that the conformal transformation changes
locally the proper time
$$ds(x)=\sqrt{g_{\mu\nu}dx^{\mu}dx^\nu} \hskip1cm \rightarrow \hskip1cm
d\tilde s(x)=\Omega(x)ds(x).$$

Since the physical phenomena should be independent of the unit chosen
locally for the length, the proper time, mass etc.  the group $C$ of
local conformal transformations should be a symmetry group of physical
laws.

In order to avoid any confusion we stress that the group $C$ has
nothing in common with the 15 parameter conformal group $SO(4,2)$
defined locally in the $M^{3,1}$by the action of Poincare, dilatation
and special conformal transformations.

Comparing the physical meaning of local conformal transformations and
the local gauge $SU(2)_L$ transformations of SM associated with the
concept of the weak isospin it seems that the conformal
transformations are not less natural symmetry transformation than the
nonabelian gauge transformations in the SM.
\medskip

We shall give now a construction of the representation of the
conformal group $C$ in the field space.  Let $\Psi$ be a tensor or
spinor field of arbitrary spin.  Define the map $$\Omega\rightarrow
U(\Omega)$$ by the formula
$$\tilde{\Psi}(x)=U(\Omega)\Psi(x)=\Omega^s(x)\Psi(x), \hskip1cm s\in R
\eqno (3.3)$$

The number $s$ is determined by the condition of conformal invariance
of field equation.  We say that field equation for $\Psi$ is conformal
invariant if there exist $s\in R$ such that $\Psi(x)$ is a solution
with the metric $g_{{\mu\nu}}(x)$ if and only if $\tilde\Psi(x)$ given
by (3.3) is a solution with the metric $\tilde{g}_{{\mu\nu}}(x)$.  The
number $s$ is called the conformal weight of $\Psi$ \cite{birel},
\cite{wald}, \cite{casta}.  It is evident that the map
$\Omega\rightarrow U(\Omega)$ defines the representation of $C$ in the
field space.

Using the above definitions one can calculate the conformal weight for
a field of arbitrary spin.  Let for instance $F_{\mu\nu}$ be the
Maxwell field on $(M^{3,1}, g)$ which satisfies the equation
$$g^{\mu\sigma}\nabla_{\sigma}F_{\mu\nu}=0 $$
$$\nabla_{[\sigma}F_{\mu\nu]}=0. $$

Using the definition of the covariant derivative $\tilde\nabla_\sigma$
with respect to $\tilde g_{\mu\nu}$ metric and (3.3) one obtains
$$\tilde{g}^{\mu\sigma}\tilde\nabla_{\sigma}(\Omega^s
F_{\mu\nu})=(n-4+s)\Omega^{s-3}
g^{\mu\sigma}F_{\mu\nu}\nabla_{\sigma}\Omega $$
$$\tilde\nabla_{[\sigma}(\Omega^s F_{\mu\nu]})=s\Omega^{s-1}
(\nabla_{[\sigma}\Omega)F_{\mu\nu]}. $$

We see that for $n\not=4$ the Maxwell equations are not conformally
invariant.  For $n=4$ the Maxwell equations are invariant if the
conformal weight $s$ equals to zero.

Similarly one can show that the Yang--Mills field strength
${F_{\mu\nu}}^a$ has the conformal weight $s=0$ whereas the massless
Dirac field has the conformal weight $s=-{3\over 2}$.  It is
noteworthy that the scalar massless field $\Phi$ satisfying the
Laplace--Beltrami equation $$\triangle\Phi=0 $$ is not conformal
invariant.  In fact it was discovered by Penrose that one has to add
to the Lagrangian on $(M^{3,1},g)$ the term $$ -{1\over 6}R\Phi^2 $$
where $R$ is the Ricci scalar, in order that the corresponding field
equation is conformal invariant with the conformal weight $s=-1$
\cite{penrose}.

\section{A unified model for strong, electro--weak and gravitational
interactions}

We postulate that the searched unified theory of strong, electro--weak
and gravitational interactions will be determined by the condition of
invariance with respect to the group $G$ $$ G=SU(3)\times
SU(2)_{L}\times U(1)\times C \eqno (4.1)$$ where $C$ is the local
conformal group defined by (3.1).  Let $\Psi$ be the collection of
vector meson, fermion and scalar fields which appear in the
conventional minimal SM for electro--weak and strong interactions.
Then the minimal natural conformal and $SU(3)\times SU(2)_{L}\times
U(1)$ --gauge invariant total lagrangian $L(\Psi)$ may be postulated
in the form:

$$ L = [L_{{G}}+L_{{F}}+L_{Y}+
L_{\Phi} + \beta\partial_\mu|\Phi|\partial^\mu|\Phi|
- {1\over 6}(1+\beta)R\Phi^{\dagger}\Phi  +
L_{{grav}} ] \sqrt{-g} \eqno (4.2)
$$

Here $L_{{G}}$ is the total lagrangian for the gauge fields
$A^{a}_{\mu}$, $W^{b}_{\mu}$ and $B_{\mu}$, $a=1,...,8$, $b=1,2,3$
associated with $SU(3)\times SU(2)_{L}\times U(1)$ gauge group
$$L_G=-{1\over 4}{F^a}_{\mu\nu}{F^a}^{\mu\nu}-
{1\over 4}{W^b}_{\mu\nu}{W^b}^{\mu\nu}-
{1\over 4}{B}_{\mu\nu}{B}^{\mu\nu}, \eqno (4.3)$$
and ${F^a}_{\mu\nu}$, ${W^b}_{\mu\nu}$ and ${B}_{\mu\nu}$ are the
conventional field strengths of gauge fields in which the ordinary
derivatives are replaced by the covariant derivatives e.g.
$${B}_{\mu\nu}=\nabla_\mu B_\nu - \nabla_\nu B_\mu, \eqno (4.4)$$
etc.; $L_{{F}}$ is the lagrangian for fermion field interacting with
the gauge fields; $L_{Y}$ represents the Yukawa interactions of
fermion and scalar fields; $L_{\Phi}$ is the lagrangian for the scalar
fields
$$L_{\Phi}=(D\Phi)^{\dagger}(D\Phi) - \lambda(\Phi^{\dagger}\Phi)^2 \eqno
(4.5)$$
where $D$ denotes the covariant derivative with connections of all
symmetry groups.  Notice that the condition of conformal invariance
does not admit the Higgs mass term $\mu^2\phi^{\dagger}\phi$ which
assures the mechanism of spontaneous symmetry breaking and mass
generation in the conventional formulation.

The term
$$\beta\partial_\mu|\Phi|\partial^\mu|\Phi| \eqno (4.6)$$
is gauge invariant. It may be surprising that (4.6) depends on
$|\Phi|$. Observe however that the lagrangian $L_\Phi$ can be written
in the form
$$(D\Phi)^{\dagger}(D\Phi)=\partial_\mu|\Phi|\partial^\mu|\Phi|
+|\Phi|^2L_\sigma(g(\Phi),W,B) \eqno (4.7)$$
where $L_\sigma(g(\Phi),W,B)$ is a gauged--sigma--model--like
lagrangian and
$$\Phi = {\phi_u\choose \phi_d}= g(\Phi){0\choose |\Phi|}, \hskip.5cm
g(\Phi)={1\over|\Phi|}\pmatrix{\bar\phi_d & \phi_u \cr
-\bar\phi_u & \phi_d} \eqno (4.8)$$
where $g(\Phi)$is $SU(2)_L$ gauge unitary matrix.

We see therefore that the term like (4.6) is already present in the
conventional $L_\Phi$ lagrangian.

The term
$$- {1\over 6}(1+\beta)R\Phi^{\dagger}\Phi \eqno (4.9)$$
is the Penrose term. which assures that the lagrangian
(4.2) is conformal invariant.

The last term in (4.2) is the Weyl term
$$ L_{{grav}} = -\rho C^2, \hskip1cm\rho>0, \eqno (4.10)$$
where $C_{\alpha\beta\gamma}^\delta$ is the Weyl tensor which is
conformally invariant.  Using the Gauss--Bonnet identity we can write
$C^2$ in the form
$$C^2=2(R^{{\mu\nu}}R_{{\mu\nu}}-{1\over 3}R^2). \eqno (4.11)$$

We see that the condition of conformal invariance does not admit in
(4.2) the conventional gravitational Einstein lagrangian
$$L=\kappa^{-2} R\sqrt{-g}, \hskip1cm \kappa^2=16\pi G. \eqno (4.12)$$

It was shown however by Stelle \cite{Stelle} that quantum gravity
sector contained in (4.2) is perturbatively renormalizable whereas the
quantum gravity defined by the Einstein lagrangian (4.12) coupled with
matter is nonrenormalizable \cite{deser}.  Hence, for a time being it
is an open question which form of gravitational interaction is more
proper on the quantum level.  We show in Section 7 that the Einstein
lagrangian (4.12) may be reproduced by Penrose term if the physical
scale is properly determined.  In Section 8 we discuss the role of
quantum effects which may reproduce the lagrangian (4.12) and give
the classical Einstein theory as the effective induced gravity.

Notice that conformal symmetry implies that all coupling constants in
the present model are dimensionless.

The theory given by (4.2) is our conformally invariant proposition
alternative to the standard Higgs--like theory with SSB.  Its new,
most important feature is the local conformal invariance.  It means
that simultaneous rescaling of all fields (including the field of
metric tensor) with a common, arbitrary, space--time dependent factor
$\Omega(x)$ taken with a proper power for each field (the conformal
weight) will leave the Lagrangian (4.2) unaffected.  The symmetry has
a clear and obvious physical meaning \cite{narlikar}, \cite{wald}.  It
changes in every point of the space--time all dimensional quantities
(lengths, masses, energy levels, etc) leaving theirs ratios unchanged.
It reflexes the deep truth of the nature that nothing except the
numbers has an independent physical meaning.

The freedom of choice of the length scale is nothing but the gauge
fixing freedom connected with the conformal symmetry group.  In the
conventional approach we define the length scale in such a way that
elementary particle masses are the same for all times and in all
places.  This will be the case when we rescale all fields with the
$x$--dependent conformal factor $\Omega(x)$ in such a manner that the
length of the rescaled scalar field doublet is fixed i.e.

$$\tilde\Phi^{\dagger}\tilde\Phi={v^2 \over 2}=const. \eqno (4.13)
$$

(We shall discuss the problem of mass generation in details in
Section 5.)

The scale fixing for the conformal group (4.13) is distinguished by
nothing but our convenience.  Obviously we can choose other gauge
fixing condition, e.g.  we can use the freedom of conformal factor to
set

$$\sqrt{-\tilde{\tilde g}}=1; \eqno (4.14)$$
this will lead to other local scales but it will leave physical
predictions unchanged.

Consider, for example the scale fixing condition (4.14).  Imposing
(4.14) on the conformal invariant theory given by (4.2) we obtain
the lagrangian $\tilde{\tilde L}(\tilde{\tilde \Psi},
\tilde{\tilde V_\mu}, \tilde{\tilde \Phi}, \tilde{\tilde g}_{\mu\nu})$
describing dynamics of the fields $\tilde{\tilde \Psi}$,
$\tilde{\tilde V_\mu}$, $\tilde{\tilde \Phi}$, $\tilde{\tilde
g}_{\mu\nu})$.  The arguments of $\tilde{\tilde L}$ stand for all
fermion, vector, scalar and tensor fields of the model and fulfill the
condition (4.14).  $\tilde{\tilde L}$ is no longer conformal invariant
as the scale was fixed by (4.14).  We can change variables of
$\tilde{\tilde L}$ according to the rule

$$\tilde \Psi=\Biggl({\sqrt{2}|\tilde{\tilde \Phi}|\over v}
\Biggr)^{-3/2}\tilde{
\tilde \Psi} \eqno (4.15a)$$
$$\tilde V_\mu=\tilde{\tilde V_\mu} \eqno (4.15b)$$
$$\tilde g_{\mu\nu}=\Biggl({\sqrt{2}|\tilde{\tilde \Phi}|\over v}
\Biggr)^2
\tilde{\tilde g}_{\mu\nu}\eqno (4.15c)$$
$$\tilde \Phi=\Biggl({\sqrt{2}|\tilde{\tilde \Phi}|\over
v}\Biggr)^{-1}\tilde{\tilde \Phi}\eqno (4.15d)$$
where $\tilde g$ is no longer restricted but $\tilde \Phi$ fulfills
(4.13) what follows from (4.15d).  Such a change of variable is an
example of conformal transformation but, as was said, it is not a
symmetry of $\tilde{\tilde L}$.  In fact we have

$$
\tilde{\tilde L}(\tilde{\tilde
\Psi}(\tilde \Psi, \tilde V_\mu, \tilde \Phi, \tilde
g_{\mu\nu}), \tilde{\tilde V_\mu}(\tilde \Psi, \tilde V_\mu,
\tilde \Phi, \tilde g_{\mu\nu})
, \tilde{\tilde \Phi}(\tilde \Psi, \tilde V_\mu, \tilde \Phi,
\tilde g_{\mu\nu})
, \tilde{\tilde g}_{\mu\nu}(\tilde \Psi, \tilde V_\mu, \tilde
\Phi, \tilde g_{\mu\nu}))=$$
$$=\tilde L(\tilde \Psi, \tilde V_\mu,
\tilde \Phi, \tilde g_{\mu\nu}) \eqno (4.16)$$
where $\tilde L(\tilde \Psi, \tilde V_\mu, \tilde \Phi, \tilde
g_{\mu\nu})$ is the lagrangian which one would obtain by imposing the
scale fixing condition (4.13) directly on (4.2). It should be stressed
that the functional form of $\tilde{\tilde L}$ in terms of its
arguments is different than $\tilde L$ of its arguments (compare  (5.1)
and (8.2) for concrete examples). In such a sense
theories obtained from different scale fixings are mathematically
equivalent.  They will be equivalent also physically if
identifications of physical and mathematical objects in the theories
being compared will be consistent with theirs mathematical
equivalence.  For example if we assume that $\tilde g$ describes
physical metric we cannot assume that this metric is described also by
$\tilde{\tilde g}$.
\medskip

\section{Generation of lepton, quark and vector boson masses}

We demonstrate now that using the conformal group scale fixing
condition
(4.13) we can generate the same lepton, quark and vector meson masses
as in the conventional SM without however use of any kind of Higgs
mechanism and SSB.

In fact inserting the scale fixing condition (4.13) into the
Lagrangian (4.2) we obtain
$$ \tilde L=L^{scaled} = [L_{{G}}+L_{F}+L_{\Phi}^{scaled}+
L_{Y}^{scaled} - {1\over 12}v^2 R  +  L_{{grav}} ] \sqrt{-g},
\eqno (5.1)
$$
in which the condition (4.13) was inserted into $L _{\Phi}$ and
$L_{Y}$.  We should use the symbol $\tilde\Phi$, $\tilde\Psi$ etc.
for the rescaled fields in (5.1), however for the sake of simplicity
we shall omit ~"~$\tilde{}$~"~ sign over fields in the following
considerations.

The condition (4.13) together with the unitary gauge fixing of
$SU(2)_{L}\times U(1)$ gauge group, reduce by (4.8) the Higgs doublet
to the form
$$\Phi^{{gauge}}={1\over\sqrt{2}}{0\choose v}, \hskip 1cm v>0
\eqno (5.2)$$
and produce the tree level mass terms for leptons, quarks and vector
bosons associated with $SU(2)_L$ gauge group.  For instance the
$\Phi$--lepton Yukawa interaction $L^l_{Y}$
$$L^l_{Y}=-\sum_{i=e,\mu,\tau}G_{i}\bar {l_{i}}_R(\Phi^\ast {l_{i}}_L)
+h.c.$$
passes into
$${L^l_{Y}}^{gauged}=-{1\over\sqrt{2}}v(G_{e}\bar
e e + G_\mu \bar\mu \mu + G_\tau \bar\tau \tau) \eqno (5.3)$$
giving the conventional, space--time independent lepton masses

$$m_{e}={1\over\sqrt{2}}G_{e}v, \hskip1cm
m_{\mu}={1\over\sqrt{2}}G_{\mu}v, \hskip1cm
m_{\tau}={1\over\sqrt{2}}G_{\tau}v. \eqno (5.4)$$

Similarly one generates from $\Phi$--quark Yukawa interaction
$L^q_{Y}$ the corresponding quark masses.  In turn from
$L_{\Phi}$-lagrangian (4.5) using the gauge condition (5.2) one
obtains
$$(D_\mu\tilde\Phi)^\dagger
D^\mu \tilde\Phi={g^2_2 v^2\over4}W_\mu^{+}W^{\mu-}+
{g^2_1+g^2_2\over8}v^2Z^2$$
where
$$Z_\mu=-\sin\theta_W B_\mu + \cos\theta_W {W^3}_\mu,
\hskip1cm \cos \theta_W={g_2\over\sqrt{g^2_1+g^2_2}}.$$

Hence one obtains the following vector mesons masses
$$m_{W}={v\over 2}g_2, \hskip1cm
m_{Z}={m_W\over \cos\theta_W}. \eqno (5.5)$$

It is remarkable that the analytical form for tree level fermion and
vector meson masses in terms of coupling constants and the parameter
$v$ is the same as in the conventional SM.  We see therefore that
the Higgs mechanism and SSB is not indispensable for the fermion and
vector mesons mass generation!

We note that the fermion--vector boson interactions in our model are
the same as in SM.  Hence analogously as in the case of conventional
formulation of SM one can deduce the tree level relation between $v$
and $G_F$ -- the four--fermion coupling constant of $\beta$--decay:
$$v^2=(2G_F)^{-1}\rightarrow v=246GeV. \eqno (5.6)$$
Here we have used the standard decomposition $g^{\mu\nu}\sqrt{-g}=
\eta^{\mu\nu}+\kappa^\prime h^{\mu\nu}$ (see e.g.  \cite{capper})
which reduces the tree level problem for the matter fields to the
ordinary flat case task.

We see therefore that the resulting expressions for masses of physical
particles are identical as in the conventional SM.

Let us stress that the scale fixing condition like (4.13) does not
break $SU(2)_{L}\times U(1)$ gauge symmetry.  The symmetry is broken
(or rather one of gauge equivalent description is fixed) when (4.13)
is combined with unitary gauge condition of electro--weak group
leading to (5.2).  However, also after imposing of a gauge
condition like (5.2) we have a remnant of both the conformal and
$SU(3)\times SU(2)_{L}\times U(1)$ initial gauge symmetries:  this is
reflected in the special, unique relations between couplings and
masses in our model

\section{Precision tests of electro--weak interactions and high
energy behavior in the present model.}

Our model represents in fact the gauge field theory model with massive
vector mesons and fermions.  It is well--known that such models are
in general nonrenormalizable \cite{sigmanonrenormalizable}.  We remind however
that in nonrenormalizable Fermi model for weak interactions we can
make a definite predictions for low energy phenomena e.g.  for $\mu$
or neutron decays.  Similarly the recent progress with so called
Generalized Equivalence Theorem allows to make definite predictions
for the scattering operator in nonrenormalizable models like gauged
nonlinear $\sigma$--model or other nonrenormalizable gauge field
theory models \cite{equivalence}.  Hence in our model we can obtain
definite predictions for electro--weak phenomena if we consider
processes with energy $\sqrt{s}$ below some ultraviolet (UV) cutoff
$\Lambda$.  We wish to demonstrate that the cutoff $\Lambda$ is
determined by the Higgs mass $m_H$ appearing in the Standard Model.
Hence, from this point of view, Higgs mass is nothing else as the UV
cutoff which assures that the truncated perturbation series is
meaningful.  We shall try to elucidate this problem on the example of
so called precision tests of electro--weak theory.

One--loop radiative corrections to various electro--weak quantities or
processes can be expressed in terms of three quantities $\Delta r$,
$\Delta\rho$ and $\Delta k^\prime$.  We refer to the recent excellent
review by Kniehl for the precise definitions of these quantities and
for their analytical expressions \cite{kniehl}.  For an illustration we
recall that the expression for W--meson mass, up to one loop order,
has the form

$$M_W={M_Z\over\sqrt{2}}\Biggl\{1+\sqrt{1+{2\sqrt{2}\pi\alpha\over M_Z
G_F(1-\Delta r)}}\Biggr\}^{{1\over2}} \eqno (6.1)$$
where
$$\alpha={1\over137.036},$$
$$G_F=1.16639\times 10^{-5}GeV^{-2}, \eqno (6.2)$$
$$M_Z=91.1899\pm 0.0044GeV$$
and $\Delta r(m_t, m_H)$ is the one loop correction to $\mu$--decay
amplitude which in Standard Model depends on top and Higgs masses.
Taking the experimental value for W--mass $M_W=80.21\pm 0.18GeV$ and
the recently reported top mass $m_t=174\pm 17GeV$ one gets from (6.1)
the central value of Higgs mass $m_H\cong1700Gev$ with the error of
several hundreds of GeV \cite{hioki}.

It was suggested by
Altarelli {\sl et.al} \cite{altarelli} to pass from $\Delta r$,
$\Delta\rho$ and $\Delta k^\prime$ to new quantities
$\varepsilon_{N1}$, $\varepsilon_{N2}$ and $\varepsilon_{N3}$ such
that $\varepsilon_{N2}$ and
$\varepsilon_{N3}$ depend on $m_t$ only logaritmically.
These parameters characterize the degree of
$SU(2)_L\times U(1)$ symmetry breaking and their numerical value
significantly different from zero would signal a "new physics"
\cite{schild}\cite{altarelli}.

If we calculate these parameters in our model in one--loop
approximation we find the specific class of Feynman diagrams with
fermion and vector boson loops which contributes to them.  Since some
vector boson loops will produce divergences, e.g.  in the case of
fermion -- massive vector boson coupling constant, one has to
introduce either the new renormalization constant or UV cutoff
$\Lambda$ which can be given by the formula \cite{schild}

$$\log{\Lambda^2\over\mu^2}={2\over 4-D}-\gamma_E+\log{4\pi} \eqno
(6.3)$$
where $\mu$ is the reference mass of dimensional regularization, $D$
is the space--time dimension and $\gamma_E$ is the Euler's constant.

One obtains the formula for $\varepsilon_{Ni}$ parameters in SM if one
adds to the class of Feynman diagrams in our model all appropriate
one--loop diagrams with Higgs internal lines.  Using the results of
\cite{schild} and \cite{schildk} one obtains

$$\varepsilon^{SM}_{N1}-\varepsilon^{CSM}_{N1}=
{3\alpha({M_Z}^2)\over16\pi {c_0}^2}
\log{({\Lambda^2\over{m_H}^2})}+O({{M_Z}^2\over{m_H}^2}
\log{({{M_H}^2\over{M_Z}^2})})$$
$$\varepsilon^{SM}_{N2}-\varepsilon^{CSM}_{N2}=
O({{M_Z}^2\over{m_H}^2} \log{({{M_H}^2\over{M_Z}^2})}) \eqno (6.4)$$
$$\varepsilon^{SM}_{N3}-\varepsilon^{CSM}_{N3}=
{\alpha({M_Z}^2)\over48\pi {s_0}^2}
\log{({\Lambda^2\over{m_H}^2})}+O({{M_Z}^2\over{m_H}^2}
\log{({{M_H}^2\over{M_Z}^2})})$$
where $CSM$ index of $\varepsilon_{Ni}$ means that the quantity
was calculated in our
Conformal Standard Model.  Here $\alpha({M_Z}^2)={1\over129}$ and
$c_0$ and $s_0$ are defined by the formula

$$s_0^2(1-s_0^2)=s_0^2c_0^2\equiv{\pi\alpha({M_Z}^2)\over
\sqrt{2}G_FM_Z^2}$$.

The above formulas indicate a role which plays in SM the very large
Higgs mass:  first the term $O({{M_Z}^2\over{m_H}^2}
\log{({{M_H}^2\over{M_Z}^2})})$ for $m_H>1TeV$can be disregarded and
second if we take the UV cutoff $\Lambda\simeq m_H$ then the
prediction for $\varepsilon_{Ni}$--parameters in the conventional SM
and our nonrenormalizable model coincide.  Thus the very large Higgs
mass preferred by the top mass $m_t=174GeV$ plays in the
conventional SM the role of UV cutoff parameter.  If the Higgs
particle will be not found then our model provides an extremely
natural frame--work for the description of electro--weak and strong
interactions at least up to TeV energies.

We would like to discuss now the problem of getting predictions from
our nonrenormalizable model for electro--weak and strong interactions
considered in the flat space--time.  Take the process $A+B\rightarrow
C+D$ in our model.  This process -- up to L--loop order -- will be
described by the corresponding Feynman diagrams with A, B, C and D
external lines and some number of internal fermion, massive vector
mesons, gluon and photon lines.  Since theory is nonrenormalizable one
has to introduce the proper UV cutoff $\Lambda$.

The problem of elaboration of an effective calculational scheme for
our model is considerably facilitated by the fact that introducing the
suitable Stueckelberger auxiliary fields we can transform our model
into the gauged nonlinear $\sigma$--model (GNL$\sigma$M) (see e.g.
\cite{schild}, \cite{sigmaSM} and the discussion in Section 9).
It is known that perturbative calculations in GNL$\sigma$M with cutoff
$\Lambda$ are well elaborated and lead to interesting physical
predictions for various processes \cite{schild}, \cite{equivalence}.

In fact it was recently shown that so called General Equivalence
Theorem (GET) holds in gauge field theories irrespectively if they are
renormalizable or nonrenormalizable \cite{equivalence}.  This
remarkable theorem can be applied in the case of SM for heavy Higgs at
high energy where
$$m_H, E\gg M_W, m_{f_i}$$
where E is the total energy and $m_{f_i}$ are lepton and quark masses
respectively.  It was shown that the leading parts coming from the
L--loop diagrams are those diagrams for which N defined as

$$N=power\; of\; m_H+power\; of\; E \eqno (6.5)$$
becomes maximal.  Using GET one relatively easily determines the
leading contribution for any L--loop in SM and obtains high energy
limit of a given scattering amplitude \cite{equivalence}.  In the case
of Higgsless nonrenormalizable gauge field theory model one introduces
cutoff $\Lambda$:  in this case at high energy limit defined by
inequalities
$$\Lambda>E\gg M_W, m_{f_i}$$
the leading diagrams are those for which
$$N=power\; of\; \Lambda+power\; of\; E \eqno (6.6)$$
is maximal.  Comparing (6.5) with (6.6) we see as in the case of the
$\varepsilon_{Ni}$--parameters that the UV cutoff $\Lambda$ in
Higgsless gauge models replaces the large mass $m_H$.  Using the
criterion (6.6) and GET one obtains the high energy limit of
scattering amplitude for various processes also in the
nonrenormalizable gauge models, like e.g.  in the Higgsless
GNL$\sigma$M \cite{equivalence}.

We see therefore that nonrenormalizability does not prevent us from
getting definite predictions for physical processes in the low or high
energy region from our model.  Consequently the nonrenormalizable
Higgsless models may be as a useful in description of experimental
data as the conventional SM.

We considered hence the general variant of our model with $\beta\not=0$
which leads to nonrenormalizable gauge field theory. However the special
case of our model with $\beta=0$ discussed in Section 8 gives a
renormalizable model for fundamental interactions.

\section{Gravity Sector}

Let us impose the scale fixing condition (4.13) on the lagrangian (4.2) and
collect all gravitational terms.  The lagrangian reads:
$$
L^{scaled}=[L^{scaled}_{matter}-{1\over12}(1+\beta)v^2R-2\rho(R^{\mu\nu}
R_{\mu\nu}- $$
$$ {1\over3}R^2)-{\lambda\over4}v^4]\sqrt{-g}
\eqno (7.1) $$
where we have selected the part $L^{scaled}_{matter}$ (describing the
matter interacting with gravity) from the remaining purely
gravitational terms.

The variation of (7.1) with respect to the metric $g^{\mu\nu}$
leads to the following classical equation of motion:
$$
\rho[-{2\over3}R_{;\mu;\nu}+2{{R_{\mu\nu}}^{;\eta}}_{;\eta}
-{2\over3}g_{\mu\nu}{R^{;\eta}}_{;\eta}- $$
$$ 4R^{\eta\lambda}R_{\mu\eta\nu\lambda}+ {4\over3}RR_{\mu\nu}+
g_{\mu\nu}(R^{\eta
\lambda}R_{\eta\lambda}-{1\over3}R^2)]+ $$
$${1\over12}(1+\beta)v^2(R_{\mu\nu}
-{1\over2}g_{\mu\nu}R)+{\lambda\over8}v^4g_{\mu\nu}={1\over2}
T_{\mu\nu}. \eqno (7.2)
$$

In the empty case $T_{\mu\nu}=0$ this equation is satisfied by all
solutions of an empty space Einstein equation with a properly chosen
cosmological constant $\Lambda$:

$$R_{\mu\nu}-{1\over2}g_{\mu\nu}R+\Lambda g_{\mu\nu}=0. \eqno
(7.3) $$
In fact (7.3) implies that
$$R_{\mu\nu}\sim g_{\mu\nu} \hskip1cm \Rightarrow \hskip1cm
R_{\mu\nu}={1\over4}Rg_{\mu\nu} \eqno (7.4) $$
and then
$$R_{\mu\nu}=\Lambda g_{\mu\nu}. \eqno (7.5) $$

Inserting (7.4) into (7.2) we find that the part proportional to
$\rho$ vanishes.  The remnant can be collected leading to the relation

$${1\over8}v^2g_{\mu\nu}({2\over3}(1+\beta)\Lambda-\lambda v^2)=0 \eqno
(7.6)$$
where the empty space condition $T_{\mu\nu}=0$ were used for the right
hand side of (7.6).

It is easy to conclude that (7.6) is satisfied when
$$\Lambda={3\over2(1+\beta)}\lambda v^2. \eqno (7.7)$$

Equation (7.7) relates $\lambda$ with a potentially observable
cosmological constant $\Lambda$.

Let us go back to the case with the matter.  Observe that the term
linear in the curvature appears in (7.1) with the coefficient
$-{1\over12}(1+\beta)v^2$.  In the case of $\beta=0$ we have the
old--fashion Standard Model minimally conformally coupled with
gravity.  In this case, in comparison with the Newtonian constant
entering to the ordinary Einstein's theory (4.12) the coefficient
$-{1\over12}v^2$ standing in front of $R$ in (7.1) has an opposite
sign and is smaller of many orders of magnitude ($v^2\kappa^2\approx
10^{-38}$).  If it would be the only purely gravitational term in the
theory it will mean that the geometry in tree approximation is
generated by the negative energy and with an extremal strength.  This
is the price we would have to pay for the positive kinetic term of
scalar fields in (4.5), for the gauge invariance and for the
renormalizability of the matter sector.

If we want to reproduce the correct gravitational sector already at
the classical level rather than preserve renormalizability of the
material sector we have to admit for nonzero $\beta$ coupling.  This
would lead us to a model which is equivalent to the nonrenormalizable
gauged nonlinear
sigma model in the material sector.  Accepting this price we can put

$$-{1\over 12}(1+\beta)v^2=\kappa^{-2} \eqno (7.8)$$
reproducing the Newtonian coupling in front of curvature $R$ in (7.1).
This would mean that $\beta\approx-10^{38}$!  Notice however that
taking the scale fixing condition (4.13) the term
$\beta\partial_\mu|\Phi|\partial^\mu|\Phi|$ vanishes.  Hence it looks
like that the only role of this term is to generate the proper value
of Newton constant in the Einstein--Hilbert tree level lagrangian
resulting from the Penrose term.  We will go back to this point in
Section 8.

\section{Towards the renormalizable theory.}

We have shown in Section 6 that the nonrenormalizability of our model
does not rise serious calculational problems within the energy range
presently accessible in experiments.  What more, if Higgs
particle will be not found then it cannot be generally excluded that
nonrenormalizability will be the indispensable feature of every
realistic particle theory model. It would mean that we are compulsed
to work with a
theory valid for a limited energy regions and
even for limited classes of phenomena. Clearly this situation is
unsatisfactory and people will always try to find
a general description scheme unifying different phenomena and
independent on the considered energy range.  Hopes for such a
universal description are usually set on renormalizable unified
models.  Being motivated by these hopes let us go back to the problem
of renormalizability of our model.

It is easy to see that nonrenormalizability of the matter part of
lagrangian (4.2) is connected with the presence of nonlinear
interaction (4.6).  To see this we can approximate (4.2) demanding
that
$$g^{\mu\nu}=\eta^{\mu\nu}.  \eqno (8.1)$$
This is a conformally flat
approximation rather than the flat approximation as we have the scale
fixing freedom, and the part of relations (8.1) can be understood as
making use of this freedom e.g.  in the form of condition (4.14).
Putting (4.14) into (4.2) we obtain
$$ \tilde{\tilde L}=L ^{unimodular}= L_{{G}}+L_{{F}}+L_{Y}+
L_{\Phi} + \beta\partial_\mu|\Phi|\partial^\mu|\Phi| - {1\over 6}
(1+\beta)R\Phi^{\dagger}\Phi  + L_{{grav}}\eqno (8.2)$$

Putting in turn (8.1) we obtain the conformally flat approximation
lagrangian $L_{cfa}$:

$$ L_{cfa} = L_{{G}}+L_{{F}}+L_{Y}+
L_{\Phi} + \beta\partial_\mu|\Phi|\partial^\mu|\Phi|
 \eqno (8.3)
$$

For $\beta=0$ this is just the renormalizable SM lagrangian (without
the negative scalar mass term $-\mu^2\Phi^\dagger\Phi$ however).

The presence of $\beta$--term was justified in Section 7 by the
condition of the proper Einsteinian limit of the theory at its
classical level.  This led us to the rather large value
$|\beta|\sim{m_{_{PLANCK}}^2\over v^2}$.  Fortunately considerations of
Section 7 showed that the predictions in particle sector of our
theory are insensitive to this huge value of $\beta$--coupling.

As we have mention in Section 7 for the case with $\beta=0$ after the
choice of physical scaling the obtained tree level Newton constant is
not correct. It was suggested by various authors that the corrected
Newton constant in quantum theories of gravity may be
obtained by inclusion of radiative corrections
\cite{zeldowich}\cite{zakharow} (see
\cite{adler} for a pedagogical introduction and \cite{buchbook} for
the recent review of the subject).
The authors of \cite{buchcqg} have discussed a wide class
of theories which contains also our model in the case of $\beta=0$.
They have shown that taking the proper values for the nonobserved
coupling
constants like $\rho$ or $\lambda$ and renormalization scale one may
generate the induced
Newton and cosmological constants with experimental values.
However this method is -- in our opinion -- incomplete since the
problem of mass values of elementary particles
in the framework in which gravitational constants were determined
was not considered. In fact the value of the Newton constant has not an
absolute meaning.
This constant disappear from the empty space Einstein equations.  In
the presence of matter the value of Newton constant can be rescaled
with simultaneous rescaling of masses and energy levels.  Thus the
value of induced Newton and cosmological constants must be
compared with the
effective masses of classical matter fields obtained within the same
level of perturbative analysis before going to the final conclusions
on induced Einstein lagrangian.
According to our knowledge the quantum gravity corrected expressions for the
effective
masses were not derived so far in the literature.

Until this problem is solved one cannot conclude that that the
renormalizable model for fundamental interactions with $\beta=0$
is physically meaningful.

\section{Discussion.}

The elementary particle physics is at present at a crossroad. We
have in fact three drastically different alternatives:

{\parindent0pt
I$^o$~~The Higgs particle exists, its mass will be experimentally determined
and will have the value predicted by the radiative corrections of SM.
This will confirm the SSB mechanism for mass generation, the validity of SM
frame--work and it will represent an extraordinary success of
quantum gauge field theory.

II$^o$~~The Higgs particle exists but its mass is considerable different
from that predicted by the radiative corrections of SM. This would signal
some kind of "New Physics" which will imply a reformulation of the
present version of SM.

III$^o$~~The Higgs particle does not exists. This will lead to a
rejection of SM with Higgs sector and it will give preference to Higgsless
models for fundamental interactions. In this situation we have
two general possibilities:

IIIA$^o$~~The physical Higgsless model is renormalizable. The example
of such model was discussed by us in Section 8.

IIIB$^o$~~The physical Higgsless model is nonrenormalizable.
It may be that the renormalizability of Quantum Gravity determined by
Einstein--Hilbert action integral coupled with matter fields is not an
"accident at work in quantum field theory" but it represents a universal
feature that physical fundamental interactions considered simultaneously
are nonrenormalizable. In this situation we are compulsed to use the
nonrenormalizable models of quantum field theory for a description
of fundamental interactions and we have to learn how to deduce predictions
for experiments from such models. Several nonrenormalizable models
for electroweak interactions were proposed like Schildknecht {\sl et al.}
model, \cite{schild},GNL$\sigma$M \cite{sigmaSM}, or
gauge field theory models with condensates \cite{condensates}.
We have presented in Section 4 a new unified  nonrenormalizable model
for fundamental interactions based on the gauge and local conformal
symmetry. }

Our model -- in spite of its nonrenormalizability -- provides the
definite predictions for low and very high energy interactions in
terms of the parameters of the model, energy $E$ and the cutoff
$\Lambda$.  The direct calculations of electro--weak parameters
$\varepsilon_{N1}$, $\varepsilon_{N2}$ and $\varepsilon_{N3}$
demonstrate that the Standard Model and the present model results
differ by the term proportional to $\log{\Lambda^2\over m_H^2}$:  thus
it looks like that the very high Higgs mass $m_H$ plays in SM the role
of the UV cutoff which in the present model may be replaced by
parameter $\Lambda$. We see therefore that the predictive power of our
model may be comparable with that of the conventional SM.

In view of the possibility that nonrenormalizable
nonabelian massive gauge field theories have to be used for a
description of
fundamental interactions it seems necessary to develop perturbative
and nonperturbative methods for extracting predictions for scattering
amplitudes and observables from such models.  In particular one should
develop
the corresponding Generelized Equivalence Theorems and determine
explicitly the
high energy behavior of cross sections in such models.  The comparison
of the obtained results with analytic formulas coming from Lipatov
calculations \cite{lipatov} would be very inspiring.  It would be also
useful to develop systematic two--loop calculus with UV cutoff
$\Lambda$ for electro--weak processes.  We plan in a near future to
present several examples of such calculations.

The present model allows to obtain the Einsteinian form of
gravitational interactions in the classical limit. It can be also
analyzed by means of effective action for induced gravity
\cite{buchbook}.

\bigskip

\leftline{ACKNOWLEDGMENTS}
\medskip

The authors are grateful to Prof.  Iwo Bia\l ynicki Birula,
Dr. B. Grz\c adkowski and Dr. M. Kalinowski for interesting
discussions and Dr. S.D. Odintsov for sending to us the recent
results of his group.

\end{document}